# Thermal versus Vacuum Magnetization in QED

Per Elmfors,[1,a] Per Liljenberg,[2,b] David Persson[3,b]
and Bo-Sture Skagerstam[4,b,c]

[a]*NORDITA, Blegdamsvej 17, DK-2100 Copenhagen Ø, Denmark*
[b]*Institute of Theoretical Physics, Chalmers University of Technology and
University of Göteborg, S-412 96 Göteborg, Sweden*
[c]*University of Kalmar, Box 905, S-391 29 Kalmar, Sweden*

**Abstract**

The magnetized relativistic Fermi and Bose gases are studied at finite temperature and density. In the case of the Fermi gas, the contribution to the magnetization from the vacuum becomes dominant for high magnetic fields, when the thermal contribution saturates. In the case of the charged Bose gas, the (paramagnetic) vacuum–magnetization becomes dominant when the gas changes from a diamagnetic to a paramagnetic behaviour. We furthermore find that the scalar–QED effective coupling constant for a weak non–zero external magnetic field is a decreasing function of the temperature.

[1]Email address: elmfors@nordita.dk.
[2]Email address: tfepl@fy.chalmers.se.
[3]Email address: tfedp@fy.chalmers.se.
[4]Email address: tfebss@fy.chalmers.se. Research supported by the Swedish National Research Council under contract no. 8244-311.

# 1 Introduction

The magnetostatic properties of the vacuum has hitherto not been studied to the same extent as the electrostatic case where the vacuum polarization has well–known screening effects. However, it is necessary to add the vacuum contribution to the magnetization in order that the average of the induced current equals the curl of the magnetization (see Ref.[1]). Then also the mean field equation follows from the minimization of the effective action with respect to the magnetic induction $B$ (see below). The vacuum magnetization becomes dominant in spinor– as well as scalar–QED, at extremely high field strengths. The magnetic fields associated with collapsed magnetic stellar objects may be as high as $B = \mathcal{O}(10^{10})$T [2] (cf. $m_e^2/e = \mathcal{O}(10^9)$T). At such field strengths the vacuum magnetization starts competing with the ordinary thermal magnetization. Here we study some aspects of a an ideal QED $e^+e^-$–plasma and a charged Bose gas at finite temperature $T$ and chemical potential $\mu$ in the presence of a static uniform magnetic field, in particularly the high field limit. The thermal Fermi and Bose gases in a high magnetic field have been studied earlier in Refs.[3, 4], but there the vacuum contribution was neglected. Since our numerical algorithms are converging faster at vanishing chemical potential, we shall mostly consider the neutral plasma, but the general behaviour is unchanged at finite density.

# 2 The Fermi Gas

The partition function $Z_f(B, T, \mu)$ for the relativistic ideal $e^+e^-$–plasma in presence of an external magnetic field $B$ in a sufficiently large quantization volume $V$, and the corresponding thermal part of the effective Lagrangian $\mathcal{L}_{\text{eff}}^{\beta,\mu}$, can be written as [5]

$$\frac{\log Z_f(B,T,\mu)}{\beta V} \equiv \mathcal{L}_{\text{eff}}^{\beta,\mu}$$
$$= \frac{eB}{2\pi^2} \sum_{n=0}^{\infty} \sum_{\lambda=1}^{2} \int_0^{\infty} dk \frac{k^2}{E_{\lambda,n}} \left( f_F^+(E_{\lambda,n}) + f_F^-(E_{\lambda,n}) \right) \quad . \quad (1)$$

Here the energy spectrum is $E_{\lambda,n} = \sqrt{k^2 + 2eB(n + \lambda - 1) + m^2}$, and $f_F^\pm(E)$ are the Fermi–Dirac equilibrium distributions

$$f_F^\pm(E) = \frac{1}{\exp[\beta(E \mp \mu)] + 1} \quad , \quad (2)$$

where $\beta$ is the inverse temperature. Separating the field independent part, we write $\mathcal{L}_{\text{eff}}^{\beta,\mu} = \mathcal{L}_0^{\beta,\mu} + \mathcal{L}_1^{\beta,\mu}$, where

$$\mathcal{L}_0^{\beta,\mu} \equiv \frac{\log Z_f(T,\mu)}{\beta V} = \frac{1}{3\pi^2} \int_{-\infty}^{\infty} d\omega \theta(\omega^2 - m^2) f_F(\omega) \left(\omega^2 - m^2\right)^{3/2} \quad . \tag{3}$$

Here $Z_f(T,\mu)$ is the partition function for the field–free ideal $e^+e^-$–plasma with particle energy $E = \sqrt{k^2 + m^2}$ and $f_F(\omega) = \theta(\omega) f_F^+(\omega) + \theta(-\omega) f_F^-(-\omega)$. Using the identity

$$\frac{\exp(-|x|)}{|x|} = \int_0^{\infty} \frac{dt}{\sqrt{2\pi t}} \exp\left(-\frac{1}{2}(x^2 t + \frac{1}{t})\right) \quad , \tag{4}$$

and expanding the distribution functions under the restriction $|\mu| < m$, we can write the field dependent part $\mathcal{L}_1^{\beta,\mu}$ for $|\mu| < m$ as

$$\mathcal{L}_1^{\beta,\mu} = \frac{1}{8\pi^2} \sum_{l \neq 0} (-1)^{l+1} \int_0^{\infty} \frac{dx}{x^3} \exp\left(-\frac{\beta^2 l^2}{4x} - m^2 x\right) \cosh(\beta l \mu) [eBx \coth(eBx) - 1] \quad . \tag{5}$$

In the case $\mu = 0$, Eq.(5) agrees with the result obtained in Ref.[6] where it was derived using thermal quantum field theory. Here we have shown that it can be directly derived from the canonical partition function. It is, however, not obvious how to generalize $\mathcal{L}_1^{\beta,\mu}$ to $|\mu| \geq m$, since then it appears to be divergent due to the presence of the infinite sum in Eq.(5). The latter problem has recently been solved and we refer to Ref.[5] for a general discussion. The effective Lagrangian density is

$$\mathcal{L}_{\text{eff}} = \mathcal{L}_0 + \mathcal{L}_1 + \mathcal{L}_{\text{eff}}^{\beta,\mu} \quad , \tag{6}$$

where the tree–level part is $\mathcal{L}_0 = -1/2\, B^2$, and $\mathcal{L}_1$ corresponds to the the well–known result [7]

$$\mathcal{L}_1 = -\frac{(eB)^2}{8\pi^2} \int_0^{\infty} \frac{dx}{x^3} \exp(-m^2 \frac{x}{eB}) \left\{ x \coth(x) - 1 - \frac{1}{3}x^2 \right\} \quad . \tag{7}$$

We have here performed the standard renormalizations leaving $eB$ invariant. Notice that Eq.(7) corresponds to a renormalized $l = 0$ term in Eq.(5). We then obtain

$$\mathcal{L}_1 + \mathcal{L}_1^{\beta,\mu} = -\frac{(eB)^2}{8\pi^2} \int_0^{\infty} \frac{dx}{x^3} \exp(-x \frac{m^2}{eB}) \vartheta_4\left[0, \exp\left(-\frac{eB\beta^2}{4x}\right)\right] \left\{ x \coth(x) - 1 - \frac{1}{3}x^2 \right\}$$

$$+ \frac{(eB)^2}{6\pi^2} \sum_{l=1}^{\infty} K_0(l\beta m)(-1)^{l+1} \quad , \tag{8}$$

where we have identified a $\vartheta_4$-function, given by $\vartheta_4[z,q] = 1 + 2\sum_{l=1}^{\infty} (-1)^l q^{l^2} \cos(2lz)$. The effective Lagrangian can be used to define an effective temperature dependent coupling constant [8]

$$\frac{1}{e_{\text{eff}}^2} = -\frac{1}{eB} \frac{\partial \mathcal{L}_{\text{eff}}}{\partial (eB)} \quad . \tag{9}$$

For weak fields, $eB \ll m^2$, and for $T \gg m$ the last term in Eq.(8) dominates and the corresponding effective coupling agrees with a conventional renormalization group calculation [5], which can be showed using the expansion $\sum_{l=1}^{\infty} K_0(xl)(-1)^{l+1} \to -1/2 \log x$ as $x \to 0$.

An external field is included by adding a term $\mathcal{L}_{\text{ext}} = \mathbf{j}_{\text{ext}} \cdot \mathbf{A}$ to $\mathcal{L}_{\text{eff}}$. Here $\mathbf{j}_{\text{ext}}$ is the external current which is independent of the dynamics of the system considered, such that $\nabla \times \mathbf{H} = \mathbf{j}_{\text{ext}}$, and $\mathbf{A}$ is the vector potential $\mathbf{B} = \nabla \times \mathbf{A}$. Neglecting a boundary term at infinity we rewrite $\mathcal{L}_{\text{ext}} = \mathbf{B} \cdot \mathbf{H}$. The mean field equation

$$B = H + M(B) \quad , \tag{10}$$

is then obtained by minimizing the effective action with respect to $B$, where we have included the vacuum contribution $M_{\text{vac}}$ in the magnetization, i.e.

$$M = M^{\beta,\mu} + M_{\text{vac}} = \frac{\partial}{\partial B}(\mathcal{L}_1^{\beta,\mu} + \mathcal{L}_1) \quad . \tag{11}$$

We find that the thermal part $M^{\beta,\mu}$ of the magnetization saturates for high fields ($\alpha = e^2/4\pi$)

$$eM^{\beta,\mu} \approx \frac{\alpha \pi}{3} T^2 \quad , \quad \sqrt{eB} > T \gg m \quad . \tag{12}$$

This was discussed in Ref.[3], but without considering the vacuum contribution

$$eM_{\text{vac}} \approx \frac{\alpha}{3\pi} eB \log(eB/m^2) \quad , \quad eB \gg m^2 \quad , \tag{13}$$

that starts to dominate when $eB \log(eB/m^2) \geq \pi^2 T^2$. We have numerically evaluated $M_{\text{vac}}$ and $M^{\beta,\mu}$ and the result is presented in Fig.1. In order to improve the convergence in the numerical calculation we have found it convenient to write Eq.(8) in the form $\int_0^{\infty} dx f(x) = \int_0^{\epsilon} dx f(x) + \int_{\epsilon}^{\infty} dx f(x)$ and to perform a modular transformation $x \to 1/x$ in the last integral. One can then choose $\epsilon = 1/\pi$ in order to obtain a numerically rapidly converging integral. As a curiosity we observe that Eq.(10) has a solution for vanishing external field $H = 0$, but a non–zero microscopic field $B$ at $eB/m^2 \approx \exp(3\pi/\alpha)$. This would mean that a spontaneous magnetic field would be generated at the Landau–ghost pole and therefore lead to a breakdown of Lorentz invariance. Perturbation theory can, however, not be naively extrapolated to such very large magnetic fields so any spontaneous vacuum magnetization cannot be concluded from the present calculation.

The magnetic susceptibility, $\chi = \partial M/\partial B$, is a measure of the fluctuations of the magnetization. If only the thermal part of the free energy is retained one would erroneously conclude that $\chi \to 0$ [3] and that there would be no fluctuations in $M$ in the large $B$ limit.

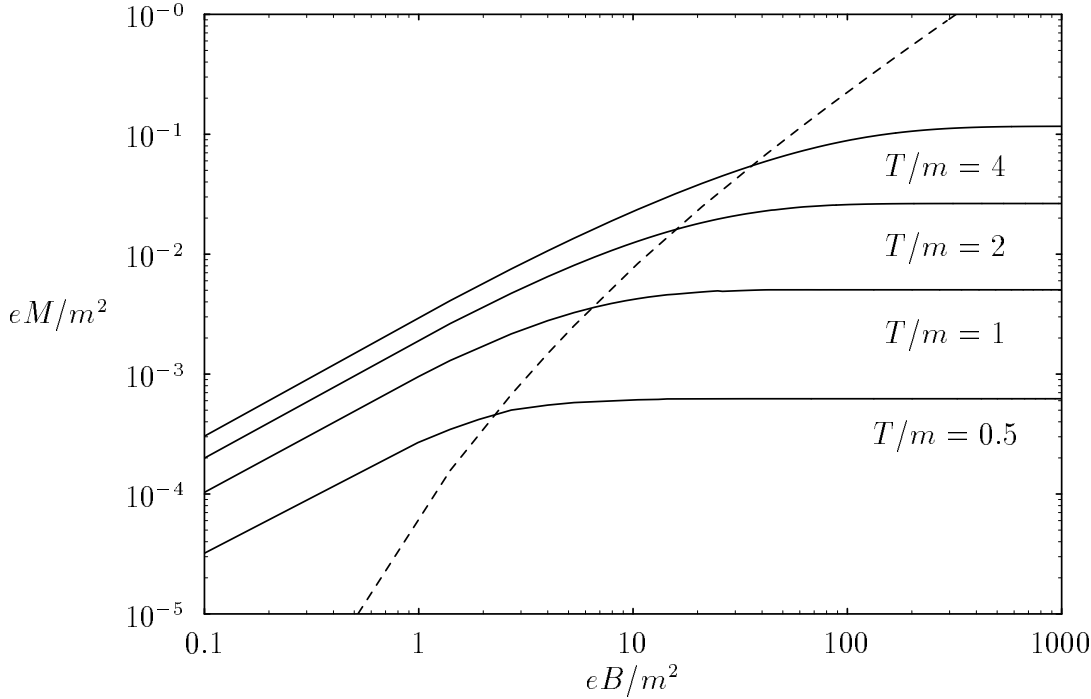

**Figure 1**: *The vacuum (dotted line) and thermal parts of the magnetization for a neutral Fermi gas. Notice that the thermal contribution saturates for large values of the magnetic field.*

## 3 The Bose Gas

The formalism used in the previous section applies also to scalar–QED. The energy spectrum is now given by $E_n = \sqrt{k_z^2 + (2n+1)eB + m^2}$ and the one–particle distributions are

$$f_B^\pm(E) = \frac{1}{\exp[\beta(E \mp \mu)] - 1} \;. \tag{14}$$

It is rather straightforward to obtain the vacuum correction [7]

$$\mathcal{L}_1 = \frac{1}{16\pi^2} \int_0^\infty \frac{ds}{s^3} \exp(-m^2 s) \left\{ \frac{eBs}{\sinh(eBs)} - 1 + \frac{(eBs)^2}{6} \right\} \;. \tag{15}$$

to the effective action. Proceeding as in the previous section we find the field independent thermal part

$$\mathcal{L}_0^{\beta,\mu} \equiv \frac{\log Z_b(T,\mu)}{\beta V} = \frac{1}{6\pi^2} \int_{-\infty}^\infty d\omega\, \theta(\omega^2 - m^2) f_B(\omega) \left(\omega^2 - m^2\right)^{3/2} \;, \tag{16}$$

where $Z_b(T,\mu)$ is now the partition function for the field independent ideal charged boson gas with particle energy $E = \sqrt{k^2 + m^2}$ and $f_B(\omega) = \theta(\omega) f_B^+(\omega) + \theta(-\omega) f_B^-(-\omega)$. In the bosonic case we always have $|\mu| \leq m$. The field dependent part of the effective action can then generally, in analogy to the previous section, be written in the form

$$\mathcal{L}_1 + \mathcal{L}_1^{\beta,\mu} = \frac{(eB)^2}{16\pi^2} \int_0^\infty \frac{dx}{x^3} \exp(-x\frac{m^2}{eB}) \vartheta_3\left[i\beta\mu, \exp\left(-\frac{eB\beta^2}{4x}\right)\right] \left\{\frac{x}{\sinh(x)} - 1 + \frac{1}{6}x^2\right\}$$
$$- \frac{(eB)^2}{24\pi^2} \sum_{l=1}^\infty K_0(l\beta m) \cosh(l\beta\mu) \quad , \tag{17}$$

where $\vartheta_3[z,q] = 1 + 2\sum_{l=1}^\infty q^{l^2} \cos(2lz)$. For weak fields, $eB \ll m^2$, and for $T \gg m$ the last term in Eq.(17) dominates and can be used to calculate the effective temperature-dependent coupling as in the previous section. One then uses the expansion $\sum_{l=1}^\infty K_0(xl) \to \pi/(2x) + 1/2 \log x$ as $x \to 0$. Even though the coefficient in front of the logarithmic term in this case agrees with an asymptotic renormalization group analysis, the term linear in $T$ leads to an effective coupling which is a *decreasing* function of the temperature. The difference between bosonic and fermionic QED, in this respect, can be understood from a one–loop calculation of the photon polarization tensor. Since we consider a static magnetic field we do not obtain any dominant thermal mass of order $e^2T^2$. For fermions the leading high $T$ term is a logarithm which is related to the UV divergence and the coefficient is the same as for the standard $\beta$–function. In the bosonic case there is an IR divergence at high $T$ from the Bose–Einstein distribution which generates a term linear in $T$. One can easily verify that the coefficient from the polarization tensor agrees with the one obtained from Eq.(17).

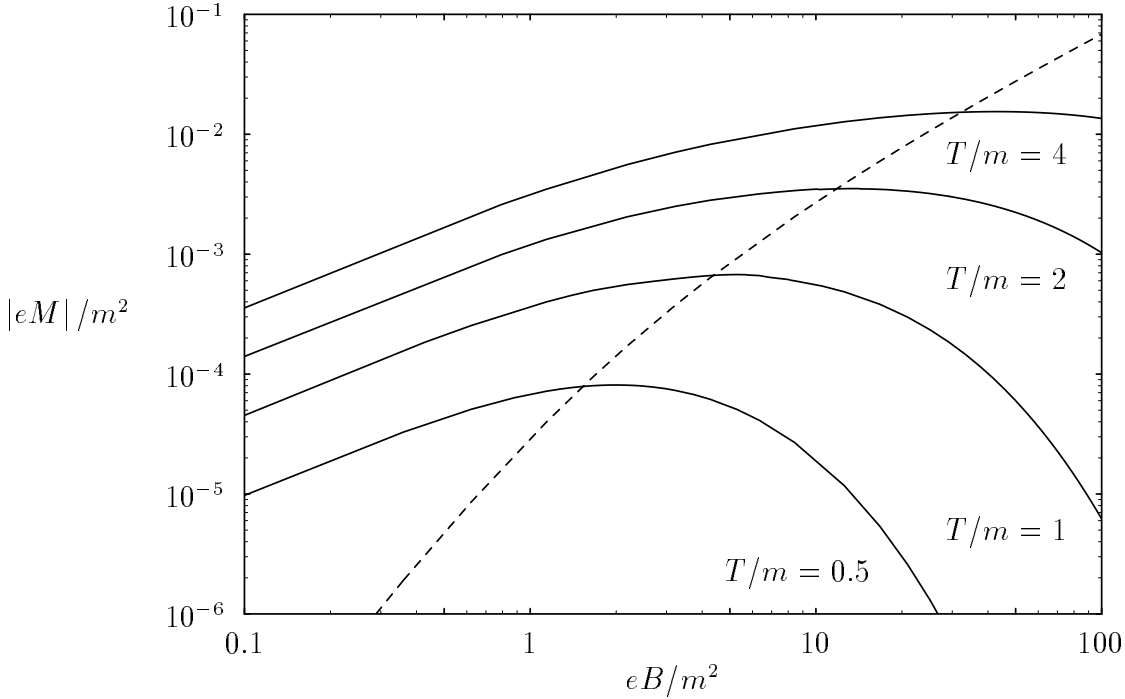

**Figure 2**: *The vacuum part (dotted line) and the modulus of the thermal part of the magnetization for a neutral Bose gas. Notice that the thermal contribution actually is negative, and that the thermal part of the magnetic susceptibility is changing sign.*

One can, furthermore, verify that the vacuum contribution $\mathcal{L}_1$ dominates over the thermal parts of the effective action if the magnetic field is sufficiently large. We have numerically evaluated $M_{\text{vac}}$ and $M^{\beta,\mu}$ and the result is presented in Fig.2. We see that the vacuum contribution starts to dominate the magnetization of the system when the gas changes from a diamagnetic to a paramagnetic behaviour.

## ACKNOWLEDGEMENT

One of the authors (B.-S. S.) thanks NFR for providing the financial support. It is a pleasure to thank the organizers of the 3rd Workshop on Thermal Field Theories, 1993, for providing a stimulating atmosphere during which parts of the present work were initialized. We also thank A. Sjölander for a stimulating discussion.